\documentclass[twocolumn,aps,prl,amsmath,amssymb,showpacs,preprintnumbers]{revtex4}
\begin{document}

\preprint{IGPG--06/11--3, AEI--2006--085}

\title{Formation and Evolution of Structure in Loop Cosmology}

\author{Martin Bojowald}
\affiliation{Institute for Gravitational Physics and Geometry,
The Pennsylvania State University, 104 Davey Lab, University Park,
PA 16802, USA}

\author{Hector H.~Hern\'andez}
\affiliation{Max-Planck-Institut f\"ur Gravitationsphysik, Albert-Einstein-Institut,\\
Am M\"uhlenberg 1, D-14476 Potsdam, Germany}

\author{Mikhail Kagan}
\affiliation{Institute for Gravitational Physics and Geometry,
The Pennsylvania State University, 104 Davey Lab, University Park,
PA 16802, USA}

\author{Parampreet Singh}
\affiliation{Institute for Gravitational Physics and Geometry,
The Pennsylvania State University, 104 Davey Lab, University Park,
PA 16802, USA}

\author{Aureliano Skirzewski}
\affiliation{Max-Planck-Institut f\"ur Gravitationsphysik, Albert-Einstein-Institut,\\
Am M\"uhlenberg 1, D-14476 Potsdam, Germany}


\newcommand{\lP}{\ell_{\mathrm P}}
\newcommand{\vp}{\varphi}
\newcommand{\vt}{\vartheta}

\newcommand{\md}{{\mathrm{d}}}
\newcommand{\tr}{\mathop{\mathrm{tr}}}
\newcommand{\sgn}{\mathop{\mathrm{sgn}}}

\newcommand*{\R}{{\mathbb R}}
\newcommand*{\N}{{\mathbb N}}
\newcommand*{\Z}{{\mathbb Z}}

\begin{abstract}
 Inhomogeneous cosmological perturbation equations are derived in loop
 quantum gravity, taking into account corrections in particular in
 gravitational parts. This provides a framework for calculating the
 evolution of modes in structure formation scenarios related to
 inflationary or bouncing models. Applications here are corrections to
 the Newton potential and to the evolution of large scale modes which
 imply non-conservation of curvature perturbations possibly noticeable
 in a running spectral index.  These effects are sensitive to
 quantization procedures and test the characteristic behavior of
 correction terms derived from quantum gravity.
\end{abstract}

\maketitle

Cosmology has provided a successful paradigm for structure formation
in our universe through an inflationary phase \cite{Guth} in early
stages. Conceptually, however, the scenario is incomplete due to the
presence of past singularities \cite{InflSing}. At such a
singularity, the classical theory of general relativity breaks down
and has to be replaced by an extended framework which remains
well-defined even at very high curvatures. Since this requires
modifications to general relativity at early stages of cosmic
evolution, there can then also be corrections to the usual scenario of
structure formation which might eventually be observable. While
dimensional arguments and low energy effective theory indicate that
effects are very small, given by the tiny ratio of the Planck length
$\ell_P=\sqrt{G\hbar}$ to the Hubble length $H^{-1}$, a detailed
analysis is required and may reveal more sizeable effects. This is
what we provide in this letter in the framework of loop quantum
gravity \cite{Rev}, a non-perturbative background independent approach
to quantize gravity.

Loop quantum gravity is one of the approaches where singularity
resolution has been investigated using loop quantum
cosmology \cite{LivRev} which results in the resolution of singularities
in various situations including inhomogeneous ones
\cite{Sing,HomCosmo,Spin,SphSymmSing}. Semiclassical bounce pictures
in special models have been described in
\cite{BounceClosed,KasnerBounce,QuantumBigBang,BouncePert}.  
A key role is played by the underlying quantum nature of spatial
geometry \cite{Area}. With such a discrete structure underlying
classical space-time, effects not captured by low energy effective
theory become possible. In particular, there are large dimensionless
parameters, such as the number of spatial lattice sites in a discrete
state, which can always spoil dimensional arguments. In such a
context, orders of magnitude of quantum corrections can only be
estimated with a detailed analysis of the effective equations arising
from quantum gravity. Suitable techniques going beyond low energy
effective theory are now available and are applied here.

On larger scales farther away from the classical singularity one can
use effective equations for the behavior of inhomogeneous
perturbations, i.e.\ equations which are of classical type but amended
by quantum correction terms as known from effective actions. Such
equations can be used to test the semiclassical viability of crucial
ingredients of quantum gravity. At the same time, they allow a
detailed study of the formation and evolution of structure including
quantum corrections.

In earlier papers \cite{PowerLoop,PowerPert}, candidate effective
equations have been used for inhomogeneous scalar fields on an
isotropic metric background.  This allowed preliminary indications,
but no systematic derivation or reliable predictions. It therefore
remained unclear which effects are to be expected.  There are, for
instance, cancellations in metric components on an isotropic
background which hide terms that may or may not be modified by quantum
gravity effects. Also, only corrections in the non-gravitational part
of the field equations were considered, while the more complicated
gravitational part plays a crucial role, too. Ignoring such
corrections does not only affect details but may even change aspects
such as the scale invariance of the expected perturbation spectrum
which is already highly constrained observationally.

For a reliable evaluation of this framework it is therefore essential
to derive full perturbation equations which take into account
corrections of gravitational dynamics. This is reported here for
scalar metric perturbations, which in Newtonian gauge are of diagonal
form
\begin{equation} \label{PertMetric}
 \delta \md s^2=-2a(\eta)^2\Phi(\eta,x)(\md\eta^2+\delta_{ab}\md x^a\md x^b)
\end{equation}
on a flat background and in conformal time $\eta$. Other gauges and
modes can be included similarly, although we do not do this here for
the sake of simplicity. Although quantum gravity is not formulated for
classical metrics, the form of the metric (\ref{PertMetric}) plays the
role of selecting the corresponding quantum regime where an effective
description is derived. This happens by picking semiclassical states
of quantum gravity which are peaked on the given class of metrics,
i.e.\ expectation values of metric operators are of the prescribed
form and fluctuations around these values are small. Expectation
values of the Hamiltonian operator in those states give the
Hamiltonian of the effective theory and thus effective equations
\cite{EffAc,Karpacz}. Although the full quantization is background
independent and non-perturbative, which is crucial for some properties
of the quantum theory such as its spatial discreteness, a cosmological
background is introduced in evaluating the theory through states and
effective equations. This puts the scenario in the usual context of
cosmological perturbation theory, albeit including quantum
corrections. Being derived from general semiclassical states which are
not Lorentz invariant unless one restricts oneself to a vacuum state,
effective equations may not be manifestly covariant even if the
underlying quantum theory is covariant \cite{Karpacz}.

Compared to the standard derivation of cosmological perturbation
equations \cite{CosmoPert}, loop quantum gravity is in a different
situation because it is based on a canonical quantization. Lagrangians
are thus not quantized directly, but Hamiltonians are used which
provides an alternative but fully equivalent classical formulation. At
this step, no new physics enters but it makes the formulation of
quantum gravity possible. To apply this to the question of interest
here requires a derivation of perturbation equations in a canonical
framework, which allows one to include effective quantum modifications
which arise for the effective Hamiltonians. Correction terms in the
evolution equations then follow uniquely, but in an indirect
manner. Thus, basic quantum modifications can have several complicated
effects in the perturbation equations which allow one to test the
underlying theory in a non-trivial manner.

For loop quantum gravity in particular, quantum Hamiltonians are
lattice operators taking into account the spatial discreteness of
quantum geometry \cite{QSDI,QSDV}. States are supported on lattices in
space, which we assume here to be regular to simplify calculations;
otherwise some coefficients can change but not by orders of
magnitude. Lattice links are labeled by quantum numbers $p_{v,I}$
correspoonding to elementary areas (centered at a vertex $v$ with
transversal direction $I$) building up space.  As seen from our final
equations, the lattice does not introduce a preferred direction
because the quantum Hamiltonian acting on the corresponding state is
direction independent. A Hamiltonian is constructed from basic
operators which are the elementary areas with eigenvalues $p_{v,I}$
and shift operators in those labels (related to curvature)
\cite{InhomLattice}.  Variables are thus discrete, associated with
lattices, and a classical geometry arises only in a continuum
limit. The size of the $p_{v,I}$ is given by the state as a multiple
of $\ell_{\rm P}^2$, which is a general parameter constructed from $G$
and $\hbar$ without input from quantum gravity. Minimum values where
non-perturbative quantum effects are significant are $p_{v,I}\approx
\ell_{\rm P}^2$, but actual values of $p_{v,I}$ in a semiclassical
state can well be larger giving rise to smaller quantum effects of the
order $\ell_{\rm P}^2/p_{v,I}$. These are the terms whose cosmological
implications we will study here.

From lattice operators one first obtains an effective Hamiltonian, by
taking an expectation value in a semiclassical state, as a function of
the lattice areas rather than of the spatial metric. Coming from a
lattice, such a function does not coincide with the classical one but
contains discretization and other effects (whose magnitudes are close
to the extrinsic curvature scale) in addition to $\ell_{\rm
P}^2/p_{v,I}$ terms. Since $p_{v,I}$ refers to the lattice size rather
than the Hubble area, corrections can be much larger than $\ell_{\rm
P}^2H^2$ as it was expected without discreteness. For $p_{v,I}\approx
\ell_{\rm P}^2$ even non-perturbative quantum effects have to be
included, but for a semiclassical geometry this cannot arise. On the
other hand, those corrections become arbitrarily small for
$p_{v,I}\to\infty$, but there is an upper limit for $p_{v,I}$ because
large $p_{v,I}$ imply large lattice sites. On length scales of
$\sqrt{p_{v,I}}$, discreteness is noticeable which must thus be much
smaller than scales probed by particle physics. Thus, a conservative
upper bound is $\ell_{\rm P}/\sqrt{p_{v,I}}\gg 10^{-15}$.  But during
inflation energy densities are much higher, up to
 $G\rho\approx 10^{-6}$
in Planck units, which requires $\ell_{\rm P}^2/p_{v,I}\gg
10^{-6}$. This also means that extrinsic curvature terms given,
through the Friedmann equation, 
by $\sqrt{Gp_{v,I}\rho}$ and thus
higher curvature corrections are small.  Although precise estimates of
correction terms require detailed constructions of semiclassical
states, the interplay of different corrections already suffices for a
rough estimate of orders of magnitude.

For regular lattices, all types of corrections can be determined
explicitly for a given Hamiltonian
\cite{QuantCorrPert}. 
The Hamiltonian itself, however, is not fixed uniquely so far but
subject to ambiguities such as the ordering of operators which do not
commute in a quantum theory. A choice of Hamiltonian, including
several ambiguity parameters, corresponds to a fixed theory which can
be tested phenomenologically. One can test precise aspects to
constrain such parameters, or consider the whole class of
possibilities allowed in the framework of loop quantum gravity and
check whether this general behavior is viable at all. At the current
stage, the second possibility is more reasonable to pursue and already
very instructive due to tight constraints on the general properties of
operators.  This gives rise to the indicated quantum corrections when
the resulting effective equations are expanded: First, spatial
discreteness implies the replacement of differential by difference
operators which, when expanded semiclassically on a background, result
in higher derivative and higher curvature corrections. Secondly,
inverse powers of metric components occur in Hamiltonians which would
classically diverge near a singularity but are modified at small
scales by quantum effects
\cite{InvScale}. The latter corrections of perturbative form
$\ell_{\rm P}^2/p_{v,I}$ are most relevant for sub-Planckian curvature
which is our present focus.

Correction functions in coefficients of the effective Hamiltonian thus
depend on the lattice areas $p_{v,I}$ rather than a continuous field
such as the spatial metric. Moreover, they depend on time only
implicitly through the time dependence of $p_{v,I}$. Qualitatively,
such a correction function $\alpha(p_{v,I})$ behaves in a way which
approaches classical behavior $\alpha=1$ for large $p_{v,I}$ but leads
to suppressions of otherwise diverging inverse powers for small
$p_{v,I}$ \cite{Ambig}. Most important for us is that any correction
function increases for very small $p_{v,I}$, reaches a peak of height
larger than one and then approaches the classical expectation
$\alpha=1$ from above in a perturbative expansion in $\ell_{\rm
P}^2/p_{v,I}$. For perturbations around an isotropic geometry, one can
express these corrections as functions of $H$ since
$p_{v,I}^{-1}\approx {\cal N}^{2/3}H^2$
for ${\cal N}$ lattice sites of volume $p_{v,I}^{3/2}$ in a Hubble
volume $H^{-3}$.
The large factor
${\cal N}^{2/3}$ thus magnifies all corrections $\ell_{\rm P}^2H^2$
expected in low energy effective theory. From the perturbative metric
(\ref{PertMetric}) it follows, on the other hand, that ${\cal
N}^{2/3}p_{v,I}(\eta)\propto a(\eta)^2(1-2\Phi(\eta,v))$ which allows
one to write all effective equations in terms of the scalar
perturbation $\Phi$.

The relevant gravitational dynamics is determined by the Hamiltonian
\cite{AshVarReell,HamPerturb}
\begin{equation} \label{Ham}
 \int\md^3x
N\epsilon_{ijk}\frac{(2\partial_{a}\Gamma_{b}^i
+\epsilon_{ilm}(\Gamma_a^l\Gamma_b^m-K_a^lK_b^m))
E^{a}_jE^{b}_k}{\sqrt{|\det E|}}
\end{equation}
expressed in basic fields $(E^a_i,K_b^j)$ which occur in loop quantum
gravity.  Here, $E^a_i$ is related to the spatial metric $q_{ab}$ by
$E^a_iE^b_i=q^{ab}\det q$ and $K_a^i$ is the canonical momentum of
$E^a_i$ (related to extrinsic curvature). The connection $\Gamma_a^i$
depends on spatial derivatives of $E^a_i$ and its inverse. The lapse
function $N$ is a free function but will be specified when choosing a
gauge.

For the effective Hamiltonian we keep only corrections for inverse
powers of metric components, disregarding higher curvature corrections
as it is adequate for sub-Planckian densities. There are two
contributions by inverse powers of the fields, the one explicit in
(\ref{Ham}) and the other in connection components. We thus have two
correction functions, $\alpha$ multiplying the whole integrand and
$\beta$ multiplying connection components. They appear in different
terms and will play quite different roles.  There are thus classes of
correction functions whose structure is determined theoretically and
whose parameters, describing their precise shape, can be restricted by
observations or other means such as internal consistency. Despite of
the non-uniqueness in parameters, crucial modifications are thus
characteristic in the general form.

As in any Hamiltonian system, the Hamiltonian generates equations of
motion. For the corrected Hamiltonian with $N=a(1-\Phi)$, corrected
perturbation equations of scalar modes in conformal time $\eta$
take the form
\cite{HamPerturb}
\begin{equation} \label{Perta}
\alpha^2 \beta \, \nabla^2 \Phi - 3 {\cal H} \dot \Phi - 3 
\left(1-\frac{\alpha' \bar p}{\alpha}\right) {\cal H}^2 \Phi  =  
-\frac{\kappa}{2} \alpha\bar p \, \delta T^0_0\,,
\end{equation}
\begin{widetext}
\begin{equation} \label{Pertb}
\ddot \Phi+ 
2\Phi \dot {\cal H}\left(1-\frac{\alpha^\prime \bar
p}{\alpha}\right)+3\dot\Phi {\cal H}\left(1 - \frac{2}{3} \frac{\alpha^\prime
\bar p}{\alpha}\right) + \frac{\alpha \beta}{3} \nabla^2
\Phi
\left(\alpha(\beta-1)- 4 \alpha^\prime \bar p\right)
+\Phi {\cal H}^2 \left(1-5 \frac{\alpha^\prime \bar p}{\alpha} +
4\left(\frac{\alpha^\prime \bar p}{\alpha}\right)^2 - 2 \frac{
\alpha^{\prime \prime} \bar p^2}{\alpha}\right) 
= \frac{\kappa}{2} \alpha\bar p \delta T^a_a \,,
\end{equation}
\end{widetext}
\begin{equation} \label{Pertc}
 \partial_a(\dot{\Phi}+{\cal H}\Phi(1-2\alpha'\bar{p}/\alpha))=
-\frac{\kappa}{2}
\bar{p} \delta T^0_a\,.
\end{equation}
A prime denotes derivatives with respect to $\bar{p}=a^2$ which is the
sum of all $p_{v,I}$, ${\cal H}=\dot{a}/a$, $\kappa=8\pi G$, and
$\delta T^a_b$ are perturbations of the stress-energy tensor.  For
classical values of the correction functions, $\alpha=\beta=1$, we
obtain the classical perturbation equations \cite{CosmoPert}, which
demonstrates the correct classical limit on very large scales of the
effective theory.  On intermediate scales, however, there are
corrections which may lead to detailed and reliable viability test of
cosmological scenarios in loop quantum gravity and in proposals for
potentially observable effects.

Our effective equations include gravitational corrections from quantum
gravity directly related to its basic discrete structures. They
suggest several applications on different scales, perhaps allowing
tests of different regimes of quantum gravity.  First, we isolate
gravitational effects by assuming an effective perfect fluid
background such that $T_{aa}=\bar{P}+\delta
P=wT_{00}=w(\bar{\rho}+\delta\rho)$ with a constant $w$ and
$T_{0a}=(\bar{\rho}+\bar{P})u_a$ with the energy density $\rho$,
pressure $P$ and velocity $u_a$. For curl-free velocity
$u_a=\partial_au$, we combine (\ref{Perta}) and (\ref{Pertc}) to a
Poisson equation
\begin{equation}
 \nabla^2 \Phi-\delta\mu(\bar{p})^2\Phi= \frac{\kappa\bar{p}}{2\alpha\beta}
(\delta\rho+3\alpha^{-1}{\cal H}(\bar{\rho}+\bar{P})u)
\end{equation}
with $\delta\mu(\bar{p})^2=3{\cal
H}^2\alpha'\bar{p}/\alpha^3\beta$. For classical values, $\delta\mu=0$
and we obtain the general relativistic Poisson equation corrected only
by a pressure term.  The correct classical limit is thus obtained as
$\alpha\to1$ for $\ell_{\rm P}\to0$. But since $\ell_{\rm P}$ is
non-zero, quantum effects always remain: to leading order we derive
Newton's potential which recieves quantum corrections.  Our derivation
of the Newton potential is alternative to that proposed in
\cite{BoundaryGraviton} and conceptually quite different. The precise
value of corrections depends on the Hubble parameter, or the
cosmological constant. This refers only to perturbations around a flat
isotropic cosmology as this is the setting in which we derived our
equations. The result on the Newton potential thus does not directly
apply to the solar system for which perturbation equations around the
Schwarzschild solution are required. They can be derived by the same
methods which are, however, technically more involved for a curved and
inhomogeneous background. From the general procedure we expect that
the Hubble parameter occuring in correction terms will 
effectively
be replaced by the solar mass.

Additional applications arise for structure formation. For the main
effect we combine Eqs.~(\ref{Perta}) and (\ref{Pertb}) to eliminate
stress-energy, still for an effective perfect fluid:
\begin{equation} \label{ddotPhi}
 \ddot{\Phi}+3(1+w+\epsilon_1){\cal H}\dot{\Phi}-
(w+\epsilon_2)\nabla^2\Phi+ \epsilon_3{\cal H}^2\Phi=0
\end{equation}
with quantum corrections $\epsilon_i$. The equation is sensitive to
the gravitational part of perturbation equations whose correction
terms, derived here for the first time, require quantum gravity. This
equation has a characteristic implication: classically, $\epsilon_i=0$
and the last term cancels exactly, but with quantum corrections
$\epsilon_3=-2\alpha''\bar{p}^2/\alpha$ is negative. These corrections
are usually small compared to the term $\nabla^2\Phi$, but can become
important for modes of small comoving wave number $k$ such that $ak\ll
H$. Such modes are outside the Hubble radius for which classically
$\Phi$ would be preserved. With quantum corrections, however, such
curvature perturbations are no longer conserved. This is seen by
solving the equation
$\ddot{\Phi}+(1+\nu)\dot{\Phi}/\eta+\epsilon_3\Phi/\eta^2=0$, with
$\nu=(5+3w)/(1+3w)$, for the behavior of large scale modes in conformal
time $\eta$, approximating $w$ and $\epsilon_3$ by a constant for the
qualitative effect: $\Phi(\eta)=\eta^{\lambda}$ with
$\lambda=-\frac{\nu}{2}\pm \frac{1}{2}
\sqrt{\nu^2-4\epsilon_3}$. Classically, there is one decaying mode and
a constant one, corresponding to conserved $\Phi$. With non-zero
$\epsilon_3$, however, the constant mode disappears, affecting the
power spectrum. Since not only the magnitude and possibly the sign of
$\epsilon_3$ but also cosmic evolution time depends on the mode, a
specific running of the spectral index can be expected. We comment
here only on the magnitude of corrections for large scale modes which
were created early in inflation: as before, an estimate for $\alpha-1$
gives $1\gg|\epsilon_3|\gg10^{-6}$. During inflation, conformal time
$\eta\propto e^{-H\tau}$ for modes currently visible on the largest
scales changes by a factor $e^{-60}$, such that the constant classical
solution is corrected by a factor $e^{-60\epsilon_3}\approx
1-10^2\epsilon_3$. On the lower end of $\epsilon_3$ this would not be
observable soon, but with lattice areas expected to be much closer to
the Planck scale, $\epsilon_3$ should be closer to one and the
magnification due to the number of $e$-foldings further enhances
quantum corrections to become potentially observable.

We have provided first effective cosmological perturbation equations
which include correction terms from quantum gravity. The scheme of the
derivation is systematic and general enough to include also other
modes and backgrounds. Unlike low energy effective theory which is
usually used to introduce quantum corrections in classical equations,
effective theory taking into account the spatial discreteness expected
from quantum gravity has revealed new effects whose magnitude can be
much larger than expected on dimensional grounds. In addition, effects
can be enlarged due to long cosmic evolution. How large precisely
corrections will be requires more detailed solutions of lattice
states, possibly involving new numerical schemes, which is accessible
in the framework now provided by loop quantum gravity.

{\bf Acknowledgements:} MB was supported by NSF grant PHY-0554771, HH
by the fellowship A/04/21572 of Deutscher Akademischer Austauschdienst
(DAAD) MK by the Center for Gravitational Wave Physics under NSF grant
PHY-01-14375 and PS by NSF grants PHY-0354932 and PHY-0456913 and the
Eberly research funds of Penn State.


\end{document}